\documentclass{iucrjournals}

\usepackage{hyperref}
\usepackage{comment}
\usepackage{amsmath}

\usepackage{lineno}
\nolinenumbers

\title{ToF-PAX-RIXS: A Time-of-Flight Spectrometer for combined ARPES and RIXS}

\author[a]{Chafic Fawaz\IUCrAufn{These two authors contributed equally.}}
\author[a,b]{Tom Lacmannn\IUCrAufn[1]{These two authors contributed equally.}}
\author[c]{Yaryna Lytvynenko}
\author[d]{Sergii Chernov}
\author[c]{Olena Tkach}
\author[a]{Siegmar Roth}
\author[e,f]{Jan O. Schunck}
\author[c]{Amon Lanz}
\author[g,h]{Jakob Dilling}
\author[g,h]{Lukas Bruckmeier}
\author[d]{Markus Scholz}
\author[d]{Frank Scholz}
\author[d]{Arne Meyer}
\author[a]{Dirk Fuchs}
\author[i]{Lara Schmieder}
\author[i]{Yiran Liu}
\author[i]{Matteo Minola}
\author[i]{Bernhard Keimer}
\author[d]{Dmytro Kutnyakhov}
\author[g,h]{Kai Rossnagel}
\author[j]{Olena Fedchenko}
\author[c]{Angela Wittmann}
\author[d]{Moritz Hoesch}
\author[c]{Gerd Sch\"onhense}
\author[a]{Matthieu Le Tacon\IUCrCemaillink{matthieu.letacon@kit.edu}}
\author[c]{Hans-Joachim Elmers\IUCrCemaillink{elmers@uni-mainz.de}}
\author[e,d,f]{Martin Beye\IUCrCemaillink{martin.beye@fysik.su.se}}

\affil[a]{Institute for Quantum Materials and Technologies, Karlsruhe Institute of Technology, 76021 Karlsruhe, Germany.}
\affil[b]{Laboratory for Quantum Magnetism, Institute of Physics, \` Ecole Polytechnique F\'ed\'erale de Lausanne (EPFL), 1015 Lausanne, Switzerland}
\affil[c]{Institut f\"ur Physik, Johannes Gutenberg-Universit\"at, Staudingerweg 7, D-55128 Mainz, Germany}
\affil[d]{Deutsches Elektronen-Synchrotron DESY, 22607 Hamburg, Germany}
\affil[e]{Department of Physics, Stockholm University, AlbaNova University Center, 106 91 Stockholm, Sweden}
\affil[f]{Wallenberg Initiative Materials Science for Sustainability, Department of Physics, Stockholm University, 106 91 Stockholm, Sweden}
\affil[g]{Institut f\"ur Experimentelle und Angewandte Physik, Christian-Albrechts-Universit\"at zu Kiel, 24098 Kiel, Germany}
\affil[h]{Ruprecht Haensel Laboratory, Deutsches Elektronen-Synchrotron DESY, 22607 Hamburg, Germany}
\affil[i]{Max Planck Institute for Solid State Research, Heisenbergstrasse 1, 70569 Stuttgart, Germany}
\affil[j]{Physikalisches Institut, Goethe-Universit\"at Frankfurt, 60438 Frankfurt am Main, Germany}

\begin{document} 
\maketitle 

\begin{synopsis}
ToF-PAX-RIXS is a newly designed instrument that combines Angle-Resolved Photoemission Spectroscopy (ARPES) based on a Time-of-Flight (ToF) detector and Resonant Inelastic X-ray Scattering (RIXS) using the Photoelectron Spectrometry for Analysis of X-rays (PAX) approach in a compact design.
\end{synopsis}

\begin{abstract}
We present ToF-PAX-RIXS, a compact instrument that combines time-of-flight momentum\nobreakdash-microscopy angle-resolved photoemission spectroscopy (ARPES) with resonant inelastic X-ray scattering (RIXS). The latter is based on photoelectron spectrometry for analysis of X-rays (PAX). The two techniques are implemented with a single multimode electron-optical column, allowing measurements to be performed sequentially on the same sample under identical experimental conditions. In ARPES mode, the instrument records the photoemission intensity as a function of kinetic energy and two in-plane momentum components. In PAX-RIXS mode, scattered soft X-ray photons are absorbed in a thin metallic converter, and the generated photoelectrons are analyzed in energy and position of emission on the converter. The electron kinetic energy retains the spectral information of the scattered photons, while the position is related to the scattering angle and therefore to the momentum transfer, enabling simultaneous access to a finite momentum-transfer interval. We describe the mechanical design, converter assembly, scattering geometry, and multimode electron optics of the instrument. Its PAX-RIXS performance is benchmarked at the Cu $L_3$ edge using CaCuO$_2$, with Au, Ag, and Pt converters. The measured spectra reproduce the principal features obtained with a conventional grating RIXS spectrometer, and deconvolution yields an effective reconstructed energy resolution better than $120$~meV under the present conditions. Finally, combined ARPES and PAX-RIXS measurements on the same Ca-doped YBa$_2$Cu$_3$O$_{6.45}$ sample demonstrate the capability of the instrument to correlate electronic structure and collective excitations within a common compact platform.
\end{abstract}


\section{Introduction}
The physics of quantum materials encompasses complex, exotic electronic behavior and emergent properties arising from multiple interacting degrees of freedom. The nontrivial behavior of these materials often originates from interactions between electrons and collective excitations such as phonons, magnons, or plasmons.
A comprehensive microscopic description of these systems therefore requires complementary experimental probes capable of resolving both the electronic structure and the associated collective excitation spectrum.

Angle-resolved photoemission spectroscopy (ARPES) and resonant inelastic X-ray scattering (RIXS) have emerged as two key spectroscopic techniques for addressing these complementary aspects of quantum materials.
ARPES provides direct access to the single-particle electronic structure and enables measurements of the energy- and momentum-dependent spectral function~\cite{damascelli2003, sobota2021}. 
RIXS, by contrast, is a photon-in--photon-out technique that probes collective charge, orbital, lattice, and magnetic excitations with elemental and orbital selectivity~\cite{kotaniRIXS2001, amentRIXS2011, Mitrano_PRX2024, degrootRIXS2024}. Combining the information obtained
from both techniques can reveal how collective excitations couple to and reshape the electronic structure. Ideally, such measurements should be performed on the same sample, under identical experimental conditions, and on the same microscopic region, thereby minimizing uncertainties associated with sample-to-sample variations, surface quality, domain structure, and experimental alignment.
\\
Despite their complementarity, ARPES and RIXS are normally implemented using fundamentally different detection concepts. ARPES analyzes photoemitted electrons, whereas conventional soft-X-ray RIXS spectrometers disperse the scattered photons using high-line-density gratings. Modern high-resolution RIXS instruments consequently require long spectrometer arms, typically extending over several meters, together with precise optomechanical alignment and dedicated detector systems \cite{brookesID322018,jarrigeSIX2018,zhouI212022, fujimori2023, ohtsuboBL132025}.
Momentum-dependent measurements are generally performed by mechanically changing the scattering geometry and recording spectra sequentially at discrete momentum-transfer values. These requirements make the direct integration of conventional RIXS and ARPES within a single compact instrument particularly challenging.

An alternative approach is photoelectron spectrometry for analysis of X-rays (PAX), in which the scattered photons are absorbed by a converter material and the generated photoelectrons are analyzed instead of the photons themselves \cite{krause1972, ebel1975,dakovskiNovelMethodResonant2017}. Because the kinetic energy distribution of the emitted electrons represents the convolution of the incident photon spectrum with the known photoemission response of the converter, the original photon-energy-loss spectrum can be reconstructed numerically. Recent work demonstrated PAX-RIXS using a solid-state converter and a hemispherical electron analyzer, including spatially resolved measurements that retain information on the momentum transfer \cite{Schunck2024}.

Building on this concept, we present ToF-PAX-RIXS, a compact spectrometer that implements PAX-RIXS in a time-of-flight momentum microscope and combines it with full-field ARPES within the same multimode electron-optical column. The system is permanently installed at beamline P04 of PETRA~III and occupies less than $4~\mathrm{m}^{2}$ of floor space. Switching between the two measurement modes is achieved by changing the electron-optical configuration and inserting or retracting a thin-film converter, without requiring a separate photon spectrometer. In ARPES mode, the instrument simultaneously records the photoemission intensity as a function of kinetic energy and two in-plane momentum components. In PAX-RIXS mode, the electron optics image the spatially resolved photoelectron distribution emitted from the converter. The electron energy encodes the energy of the scattered photon, while the position on the converter is related to the scattering angle and hence to the momentum transfer. The imaging concept therefore provides access to a finite momentum-transfer interval without mechanically scanning a conventional spectrometer arm.

In this article, we describe the mechanical design, sample and converter manipulation system, converter architecture, scattering geometry, and multimode electron optics of the instrument. We compare Au, Ag, and Pt converter materials and discuss the trade-off between conversion efficiency and achievable energy resolution. We then benchmark the PAX-RIXS performance at the Cu $L_{3}$ edge using CaCuO$_2$ and compare the reconstructed spectra directly with measurements obtained using a conventional grating RIXS spectrometer. Finally, we demonstrate the combined operation of the instrument by performing ARPES and PAX-RIXS measurements on the same cleaved Ca-doped YBa$_2$Cu$_3$O$_{6.45}$ sample. These results establish the feasibility of a compact, unified platform for correlating electronic structure and collective excitations using complementary electron- and photon-based spectroscopies.

\section{Operating principle of the combined ARPES and PAX-RIXS instrument}

\subsection{Time-of-flight momentum-microscopy ARPES}
In ARPES mode, photoelectrons emitted from the sample are collected by the cathode-lens objective and projected onto a delay-line detector through a time-of-flight electron-optical column. The detector impact coordinates encode the two parallel momentum components, $k_x$ and $k_y$, while the arrival time determines the electron kinetic energy. The instrument therefore records the three-dimensional photoemission intensity distribution $I(E_{\mathrm{kin}},k_x,k_y)$ in parallel \cite{schonhense2015,chernov2015,medjanikDirect3DMapping2017a, matsuiPhotoelectronMomentumMicroscope2020 , schoenhe2022,tkachMultimodeObjectiveLens2026}. 
By changing the electron-optical lens voltages, the same column can also be operated in a real-space imaging mode to record the photoemission intensity as a function of kinetic energy and two spatial coordinates $I(E_{kin}, x, y)$. This ability to switch between momentum and real-space imaging provides the basis for combining ARPES with the PAX-RIXS detection scheme described below.

\subsection{RIXS detection and instrumental integration}
RIXS determines the energy and momentum transferred from the incident photon to the sample by analyzing the inelastically scattered photons. 
The incident photon energy is tuned to a specific element and orbital resonance, thereby creating a core hole and enhancing the cross section of collective excitations \cite{kotaniRIXS2001, amentRIXS2011}. Analyzing scattered photon polarisation can also allow to study the symmetry properties of these excitations \cite{braicovich1999, amentRIXS2011, kotaniRIXS2005, nag2020, elnaggar2023, li2023, Mitrano_PRX2024}. Due to the large momentum carried by the X-rays, RIXS allows us to measure the dispersion of these collective excitations across the reciprocal space.
Conventional soft-X-ray RIXS spectrometers disperse the outgoing photons using a spatially variable high-groove-density grating and image the resulting spectrum onto a position-sensitive detector \cite{brookesID322018,jarrigeSIX2018,zhouI212022}. High resolving power typically requires an optical path of several meters in length, while momentum-dependent measurements are performed by changing the scattering geometry or rotating the spectrometer arm. 
Two decades into the development of this technique, energy resolution has improved by two orders of magnitude \cite{ichikawaRXE1996, brookesID322018}, but is now approaching technological limitations \cite{miyawaki2022, miyawakiUltrahighResolutionHigh2025}.
Such modern spectrometer setups are large and highly optimized for the purpose of achieving high energy resolution with dispersion gratings. The combination with ARPES measurements in a compact setup is thus virtually impossible.

\subsection{PAX-RIXS operating principle}
\begin{figure}[ht] 
\begin{center}
\includegraphics[width=1\textwidth]{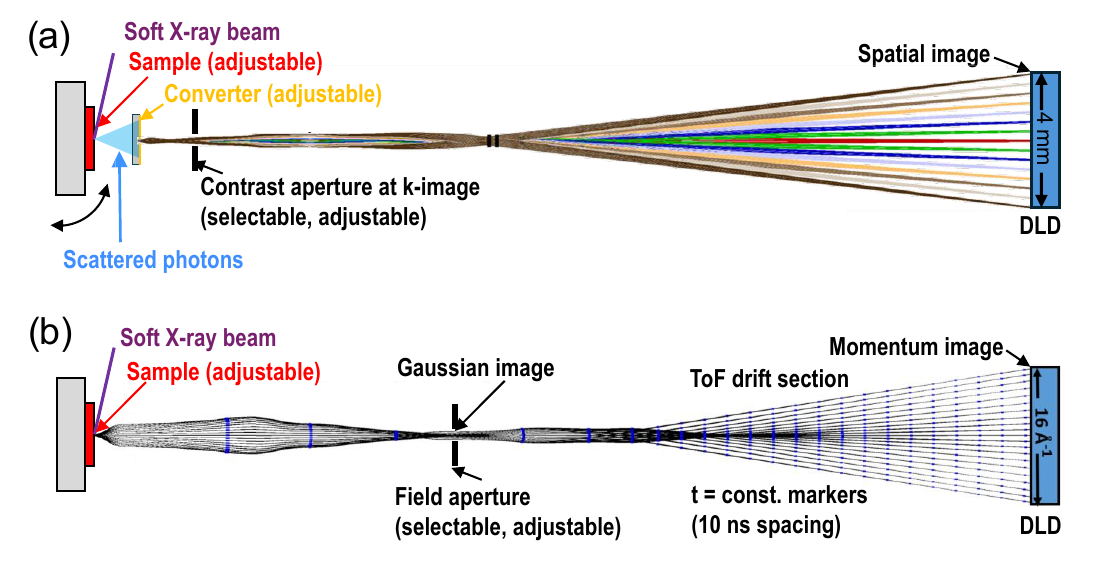} 
\end{center}
\caption{Sketch of the dual-use detector system. (a) PAX-RIXS mode, where the electron optics are set to PEEM mode, imaging the converter surface in a large field of view of several mm in diameter. The sample and converter are shown on a larger scale. (b) Momentum microscopy mode, where the converter has been removed and the electron optics image direct photoelectrons after refocusing the objective lens.}
\label{MM_PEEM}
\end{figure}

The present instrument circumvents this incompatibility by indirectly analyzing the scattered photons through photoelectron spectrometry for analysis of X-rays (PAX) \cite{krause1972,ebel1975,dakovskiNovelMethodResonant2017, Schunck2024}. In PAX-RIXS mode, a thin-film converter is inserted into the path of the photons scattered from the sample (Fig.~\ref{MM_PEEM}(a)). Absorption of scattered photons produces photoelectrons whose kinetic energy distribution is determined by the photon-energy distribution and by the photoemission response of the converter material.
These photoelectrons are collected from a large area on the converter, several mm in diameter.

The idea of using photoemission detection to analyze inelastically scattered photons was first proposed by M. O. Krause and named ‘photoelectron spectrometry for analysis of X-rays (PAX)~\cite{krause1972} by using a gas converter, and later by~\citeasnoun{ebel1975} using a solid converter, combined with numerical deconvolution of the converter signal, as well as by~\citeasnoun{benka1982}.
Recently, it was presented by~\citeasnoun{dakovskiNovelMethodResonant2017} and subsequently in our previous work~\cite{Schunck2024}, using a state-of-the-art hemispherical analyzer.

The measured spectrum is called a PAX-RIXS spectrum and contains the spectral information of the inelastically scattered photons, convolved by both the intrinsic photoemission response of the converter and the energy resolution of the time-of-flight electron analyzer. To recover the underlying RIXS spectrum, the measured signal can therefore be deconvolved using an experimentally determined response function. This response function is obtained from a reference measurement of elastically scattered photons (typically from polycrystalline graphite), for which the energy-loss distribution is negligible, recorded with the same converter and electron-optical settings.

Switching from ARPES to PAX-RIXS requires only the insertion of the converter and a change from momentum-imaging to real-space operation of the electron column. In ARPES mode (Fig.~\ref{MM_PEEM}(b)), the converter is retracted, and the electron optics directly analyzes photoelectrons from the sample. In PAX-RIXS mode, the converter is inserted, and the same electron-optical column images the photoelectrons emitted from its surface. The two measurements can therefore be carried out on the same sample without transfer to a separate spectrometer.

\subsection{Parallel encoding of momentum transfer}

In addition to the energy information encoded in the electron kinetic energy, the photoelectron emission position on the converter retains information about the emission direction of the scattered photon. For a calibrated sample--converter geometry, each converter coordinate $(x,y)$ corresponds to an outgoing photon direction $\mathbf{k}_{\mathrm{out}}$ and hence to a momentum transfer $\mathbf{Q}=\mathbf{k}_{\mathrm{in}}-\mathbf{k}_{\mathrm{out}}.$

Real-space imaging of the converter therefore permits simultaneous collection of a finite range of scattering angles and momentum transfers. This energy-and-position-resolved detection distinguishes the present implementation from earlier PAX-RIXS approaches based on non-imaging electron analyzers and is the central operating principle of the ToF-PAX-RIXS instrument.

\section{Instrument design}
\begin{figure}[ht] 
\begin{center}
\includegraphics[width=0.7\textwidth]{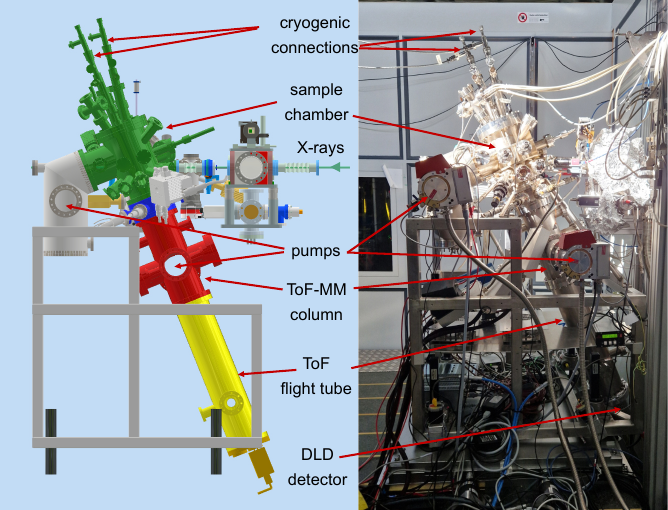} 
\end{center}
\caption{(Left) CAD overview of the assembled ToF-PAX-RIXS spectrometer. (Right) The set-up installed on beamline P04 at DESY.}
\label{Instrument_design}
\end{figure}

The ToF-PAX-RIXS instrument was designed as a very compact, permanently installed endstation that can be operated without obstructing the transmitted X-ray beam when it is not in use. The complete system occupies less than $4~\mathrm{m}^{2}$ of floor space and is installed at beamline P04 of PETRA~III at DESY. The instrument comprises a sample chamber, a multimode electron-optical column, a time-of-flight drift section,
and a delay-line detector, as shown in  Fig.~\ref{Instrument_design}.

The time structure of PETRA~III in the 40-bunch operating mode is particularly well suited to time-of-flight detection. The 192~ns bunch separation provides a sufficient time window for recording electron arrival times without temporal overlap between successive photon pulses. This is ideally suited for ToF energy recording and perfectly matches the delay-line detector's count-rate capability of 5 million counts per second.
The beamline P04 provides a beam spot of 20 × 20 $\mu m^2$ and features a plane-grating monochromator (PGM) capable of delivering a 20 meV energy bandwidth at 800 eV photon energy, while maintaining a very high photon flux of approximately $10^{12}$ photons per second~\cite{viefhaus2013}.

The sample chamber accommodates two independently operated assemblies: the sample manipulator and the converter manipulator. Both are equipped with dedicated cryogenic, mechanical, and thermal connections, allowing their temperatures and positions to be controlled independently, while preserving the fixed electron-optical axis of the momentum microscope. The independent cryogenic operation is particularly important in PAX-RIXS mode, because the spectral response of a converter based on a Fermi edge depends strongly on its temperature. A temperature sensor mounted close to the sample provides an estimate of the sample temperature during operation. Separate turbomolecular pumps connected to the sample chamber and electron-optical column maintain ultra-high-vacuum conditions during both ARPES and PAX-RIXS operation.

\subsection{Sample Chamber and Incoupling}

\begin{figure}[t] 
\begin{center}
\includegraphics[width=1\textwidth]{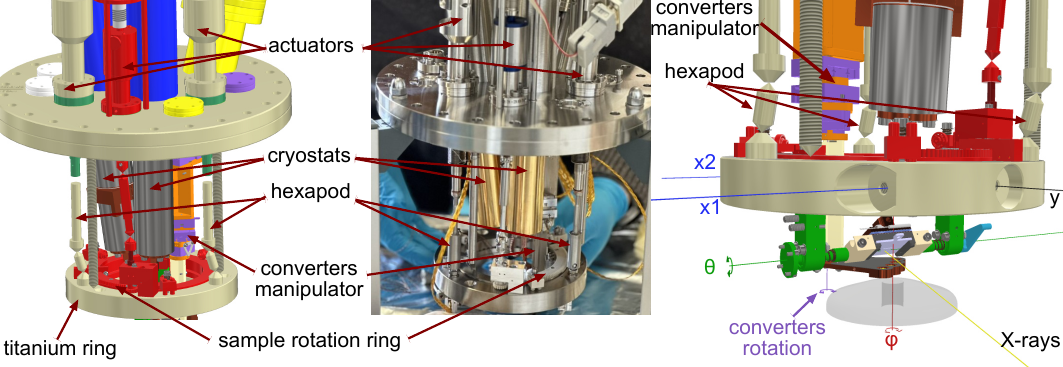}
\end{center}
\caption{(Left) CAD overview and (middle) the realized setup of the sample chamber. (Right) Translation and rotation degrees of freedom of the sample and converters.}
\label{sample_chamber_in}
\end{figure}

The sample chamber overview in Figure~\ref{sample_chamber_in} compares the mechanical design with the realized assembly and summarizes the available translational and rotational degrees of freedom. The motion degrees of freedom of both the sample and the converter are shown in Fig.~\ref{sample_chamber_in} (right~panels), as explained in the following section.

Just upstream of the sample chamber and independently supported, an incoupling chamber provides the connection to the X-ray beamline. This houses vacuum pumps for optimized sample conditions and a mirror with a central hole. The mirror allows viewing of the sample surface along the same axis as the impinging X-ray beam, easing alignment of sample spots in the X-ray beam.

\subsubsection{Sample manipulator}

The sample manipulator provides three translational and two rotational degrees of freedom. The translational motion is generated by three actuators supporting a titanium ring. Simultaneous motion of the three actuators produces translation along the electron-optical $z$ axis, whereas differential motion provides a slight tilting angle. In the notation of Fig.~\ref{sample_chamber_in}, translation along $x$ is controlled by actuators $x_1$ and $x_2$ (as indicated by the two blue axes in Fig.~\ref{sample_chamber_in} (right panel)), while translation along $y$ is controlled by the third actuator (black axis).

Two rotations determine the sample orientation relative to the incident beam and the electron-optical axis. The azimuthal angle $\varphi$, corresponding to rotation about the electron-optical axis, is controlled by an inner rotation ring mounted within the titanium support (red ring and the red axis in Fig.~\ref{sample_chamber_in} (right panel). 
The incidence angle $\theta$, defined here as the angle between the incident X-ray beam and the sample surface, is adjusted by rotation about the green axis in Fig.~\ref{sample_chamber_in} (right panel).
Together, these motions allow the illuminated sample region and the photoelectron emission geometry to be aligned independently.

\subsubsection{Converter manipulator}
The converter manipulator consists of a rotatable arm supporting up to four converter elements. Rotation of the arm selects the desired converter or alignment grid and positions it on the optical axis (indicated by the purple rotation axis in Fig.~\ref{sample_chamber_in} (right panel)). Additional translations perpendicular to the optical axis allow the selected converter to be centered relative to the entrance of the electron-optical column and to the scattered-photon footprint.

\begin{figure}[ht] 
\begin{center}
\includegraphics[width=0.35\textwidth]{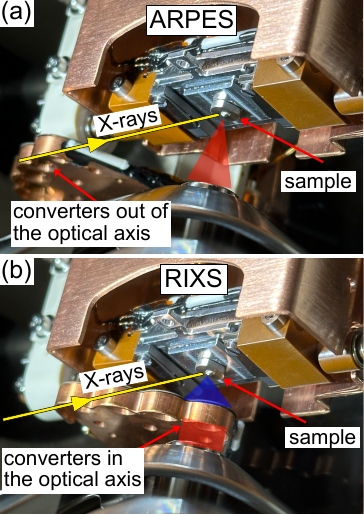} 
\end{center}
\caption{ToF-PAX-RIXS measurement configurations. (a) ARPES configuration, where the red cone represents the photo-emitted electrons. (b) RIXS configuration, where the blue cone represents the inelastic photons, and the red rectangle represents the photo-emitted electrons converted from these inelastic photons, allowing us to measure the PAX-RIXS spectrum.}
\label{ARPES_vs_RIXS_mode}
\end{figure}

The manipulator enables rapid switching between the two operating configurations shown in  Fig.~\ref{ARPES_vs_RIXS_mode}. 
In ARPES mode (Fig.~\ref{ARPES_vs_RIXS_mode}(a)), the converter arm is retracted from the electron-optical axis so that photoelectrons emitted directly from the sample enter the momentum microscope. In PAX-RIXS mode (Fig.~\ref{ARPES_vs_RIXS_mode}(b)), the selected converter is inserted on the optical axis, where it intercepts photons scattered from the sample. Photoelectrons generated at the converter are then imaged by the electron-optical column. The sample itself does not need to be removed or transferred when switching between the two modes.

\subsubsection{Cryogenic cooling system}
Cooling of both the cryogenic cold fingers for the sample and the converter assembly is a closed-cycle cooler operated with Helium gas. This two-stage and dual-branch cooler was commercially procured and retrofitted onto the housings of both flow cryostats. It consists of a central Griffith McMahon cooler (GM cooler, first stage) that cools the continuous Helium gas flow of the second stage. The second-stage circuit, driven by a separate Helium gas compressor called the recirculator, splits into branches at the GM cooler. Each branch transports the cold He gas at about 40 K to the cold fingers with additional cooling by counterflow. Inside the cold finger, a Joule-Thomson capillary assembly acts to further cool the Helium gas to near-liquid conditions and thus extracts heat from each cold finger. After streaming back, the Helium flow of the two circuits is reunited and fed into the recirculators. In each cold finger, a resistive heater and a master temperature sensor are located as well. During standard operation, the converter is kept at its lowest temperature, with a measured 8 K in the cold finger, while the sample cold finger can be cooled to a base temperature of 4.5 K. Controlled heating is then used in the sample cold finger when the sample requires an elevated temperature. The cold finger inserts can also be replaced by the original Helium flow inserts in case operation with liquid Helium is desired.

\subsection{Converters}

\subsubsection{Converter assembly}
\label{converter_assembly}

The converter assembly is the central detection element of the PAX-RIXS operating mode. It converts the scattered soft-X-ray photons into photoelectrons. The converter properties are crucial in determining the performance of PAX-RIXS, since the photoionization cross-section determines the detection efficiency and the converter spectral response is one of the principal contributions to the energy resolution of the instrument. Converter materials with narrow and simple photoemission features are advantageous for reconstructing the underlying RIXS spectrum.

A solid-state converter is particularly suitable for integration into an ultra-high-vacuum instrument. In comparison with a gas-phase converter~\cite{ebel1975}, it can be inserted and retracted mechanically and allows straightforward switching between PAX-RIXS and ARPES operation without compromising the vacuum conditions. The converter design builds on the approach introduced in Ref.~\cite{Schunck2024}.

\begin{figure}[ht] 
\begin{center}
\includegraphics[width=0.45\textwidth]{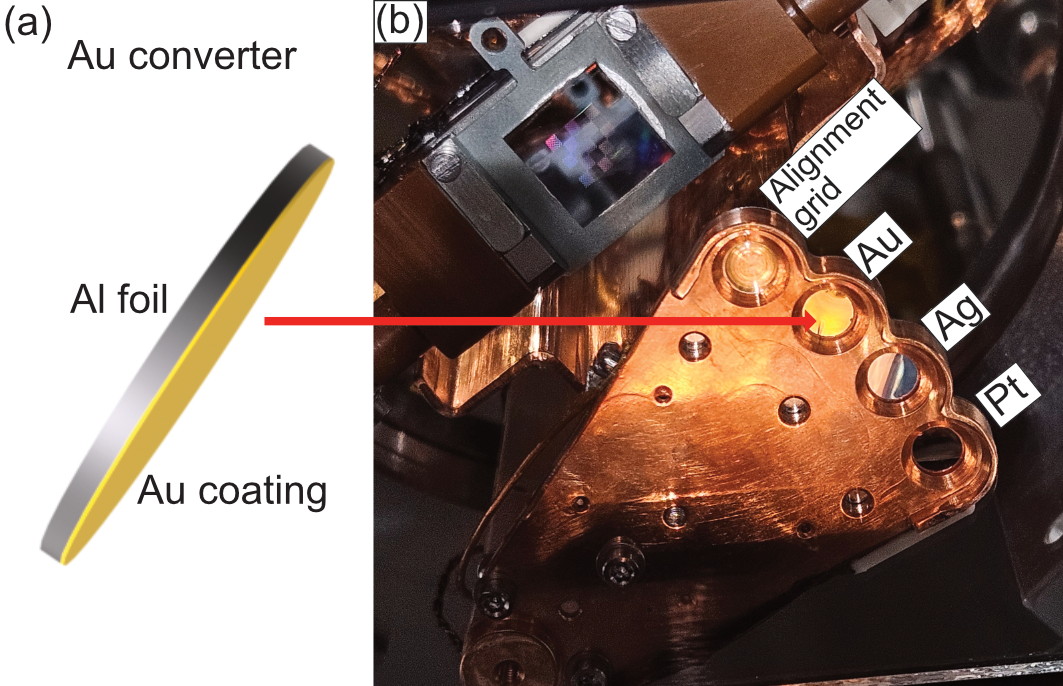} 
\end{center}
\caption{Converter assembly. (a) Schematic of the Au converter, consisting of a thin aluminum foil coated with an ultrathin gold layer as the conversion material. (b) Converter assembly holding different converters and an alignment grid.}
\label{conv_assembly}
\end{figure}

Each converter consists of an approximately 200~nm-thick aluminum support foil, which suppresses the transmission of electrons originating from the sample while remaining sufficiently transparent to soft X-rays. A thin conversion layer, typically 10--30~nm thick, is deposited onto a buffer layer on the downstream side of the foil (Fig.~\ref{conv_assembly}(a)). The converter membrane is mounted on a copper support ring with an aperture diameter of 4~mm, with the conversion layer facing away from the sample. The support arm can accommodate three converter materials together with an alignment grid, as illustrated in Fig.~\ref{conv_assembly}(b). In the present configuration, Au, Ag, and Pt converters are available.

\subsubsection{Choice of converter material}
\label{converter_material_choice}
The choice of converter material involves a trade-off between photon-to-electron conversion efficiency and spectral sharpness. The conversion efficiency depends on the incident photon energy and the relevant photoionization cross section, whereas the spectral response is determined by the intrinsic line shape of the selected photoemission feature, its thermal broadening, and the resolution of the electron analyzer. Since the converter response enters the numerical reconstruction of the RIXS spectrum, narrow and well-characterized photoemission features improve the stability of the deconvolution.

In principle, broadening contributions associated with the beamline, converter, and electron analyzer can be incorporated into the experimentally measured response function. During measurement setup, the X-ray bandwidth and spectrometer resolution can be tuned, and flux or transmission can be traded against resolution; the converter contribution is set by selecting the spectroscopic feature, as discussed below. Deconvolution may then recover spectral features that are narrower than the intrinsic width of an individual converter core level or even the energetic width settings of the beamline and instrument. In practice, however, the reconstruction becomes increasingly sensitive to statistical noise, calibration errors, and regularization parameters, and the deconvolution can generate artifacts. A converter with a sharper intrinsic response therefore reduces the demands placed on both the data quality and the reconstruction algorithm.

Silver is used as the principal converter material because the Ag~$3d$ core levels provide a favorable compromise between conversion efficiency and spectral width. At photon energies around 500~eV, the conversion efficiency can reach approximately 10\%, while the Ag~$3d$ line has a reported intrinsic full width at half maximum (FWHM) of approximately 233~meV~\cite{panaccione2005}. The large spin-orbit splitting also reduces overlap between the two core-level components, leading to better reconstruction of the deconvolved spectra. Gold provides a comparable conversion efficiency, but the Au~$4f$ core levels have a larger reported intrinsic linewidth of approximately 335~meV~\cite{takata2005}.

Platinum offers an alternative route based on the Fermi edge rather than a core-level line. At sufficiently low temperature, the Fermi edge can provide a much sharper response than the Ag~$3d$ or Au~$4f$ levels, with the thermal broadening set primarily by the converter temperature and the experimental energy resolution. This potential gain in spectral sharpness is offset by a substantially lower conversion efficiency, reported to be approximately two orders of magnitude below that of the core-level-based converters~\cite{higley2022}, together with reduced count rates at higher energy resolution. Measurements using a Pt converter therefore require longer acquisition times and correspondingly higher statistical quality.

\subsection{Converter maps the RIXS dispersion in parallel}
\label{converter_maps}

\begin{figure}[ht] 
\begin{center}
\includegraphics[width=1\textwidth]{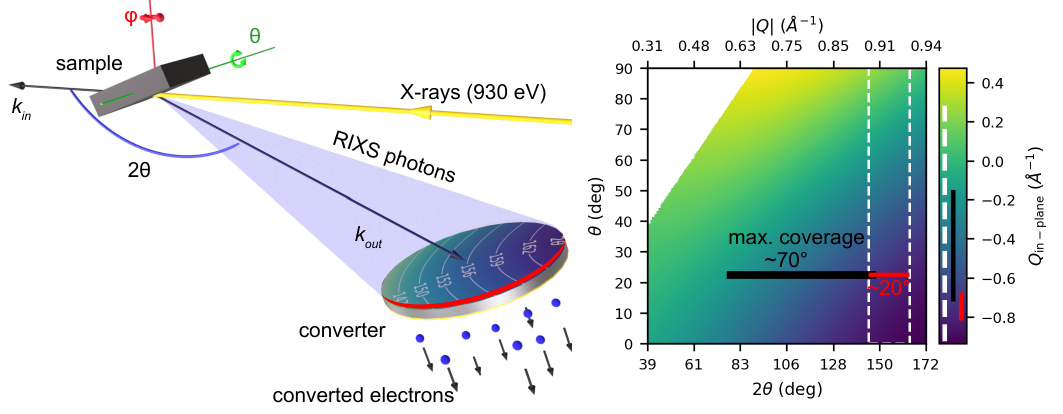}
\end{center}
\caption{(Left) Schematic of the RIXS geometry illustrating the sample rotation angles ($\theta$ and $\varphi$), and the scattering angle $2\theta$. The blue cone represents the inelastically scattered RIXS photons, which are converted into photoelectrons by the converter. Its footprint on the converter surface illustrates the finite range of $2\theta$.
(Right) Accessible scattering angle $2\theta$ (bottom axis) and corresponding transferred momentum $Q$ (top axis) at the Cu $L_3$ edge. The color scale represents the in-plane momentum transfer component $Q_{in-plane}$. The red lines denote the $2\theta$ and $Q$ ranges that are simultaneously accessible for the scattering geometry shown in the left panel highlighted by the red crescent. The red line on the color bar indicates the corresponding accessible $Q_{in-plane}$ range. The white contour and the corresponding line in the color bar indicate the corresponding $Q$, $Q_{in-plane}$ obtained by $\theta$ rotation, for the same geometry. The black line denotes the maximum $2\theta$ that can be detected simultaneously, together with the associated $Q$ and $Q_{in-plane}$ with $\theta$ rotation.}
\label{RIXS_geometry}

\end{figure}

In a conventional RIXS spectrometer, the scattering angle  $2\theta$ is selected by the position of the spectrometer arm. In the present instrument, the outgoing photons are collected over a finite solid angle by the extended converter surface (Fig.~\ref{RIXS_geometry}). Different positions on the converter therefore correspond to different outgoing photon directions and, consequently, to different momentum transfers.

For an incident photon wave vector $\mathbf{k}_{\mathrm{in}}$ and an outgoing wave vector $\mathbf{k}_{\mathrm{out}}$, the transferred momentum is
$\mathbf{Q}=\mathbf{k}_{\mathrm{in}}-\mathbf{k}_{\mathrm{out}}$. Since the photon-energy loss is small compared with the incident photon energy in the present soft-X-ray experiments, the magnitudes of the incident and outgoing wave vectors may be approximated as equal. The magnitude of the momentum transfer is then determined by the scattering angle $2\theta$ according to $Q=2k\sin(2\theta/2)$. 

Figure~\ref{RIXS_geometry} (right panel) illustrates the correspondence between converter position, scattering angle, and momentum transfer. For the geometry shown in the left panel, the illuminated converter area spans an angular interval of approximately $20^\circ$. This interval is recorded simultaneously and therefore provides parallel access to a finite range of momentum transfer without moving either the sample or the converter. A larger fraction of the converter surface can in principle be imaged, corresponding to an angular acceptance of up to approximately $70^\circ$, although the practically usable interval is also limited by the converter illumination, electron-optical field of view, spatial resolution, and counting statistics. Different ranges in $2\theta$ are sampled by translating the sample along the X-ray beam direction.
Rotation of the sample changes the projection of $\mathbf{Q}$ onto the crystallographic directions and extends the accessible in-plane momentum range. The specific values of $Q$ and $Q_{\mathrm{in-plane}}$ shown in Fig.~\ref{RIXS_geometry} were calculated for a photon energy of approximately 930~eV (at the Cu $L_3$ edge) and for the sample orientation indicated in the figure.
The geometry shown in the left panel of Fig.~\ref{RIXS_geometry} provides simultaneous access to a scattering-angle interval of approximately $20^{\circ}$, centered around $2\theta \approx 156^{\circ}$. At the Cu $L_3$ edge, this corresponds to a momentum-transfer range of $Q \approx 0.90$--$0.93~\mathrm{\AA}^{-1}$. For a fixed sample rotation angle of $\theta = 22.5^{\circ}$, the corresponding in-plane component ranges from $Q_{\mathrm{in\mbox{-}plane}} \approx -0.81$ to $-0.68~\mathrm{\AA}^{-1}$, as indicated by the red crescent in the left panel of Fig.~\ref{RIXS_geometry} and by the red line in the right panel.
This is equivalent to approximately $-0.49$ to $-0.42$ reciprocal lattice units (r.l.u.) for the lattice parameter of a high-temperature superconducting cuprate considered here as an example.

\section{Electron optics}
The electron-optical column supports three distinct operating modes: X-ray photoemission electron microscopy (X-PEEM) for real-space imaging of the sample, momentum microscopy for ARPES, and large-field-of-view X-PEEM imaging of the converter in PAX-RIXS mode. Ideally, switching between these modes should not require repositioning the sample, so that the same sample region remains under investigation.

These requirements differ substantially from those of conventional PEEM and momentum-microscope instruments. In the X-PEEM and ARPES configurations, the converter is retracted and the sample--front-lens distance exceeds 10~mm, considerably larger than the typical 2--4~mm cathode-lens gap. In PAX-RIXS mode, the electron optics must instead image a converter area with a diameter of up to 4~mm, which is substantially larger than
the field of view normally accessible in low-energy PEEM operation.

To satisfy these requirements, systematic electron-ray-tracing calculations were performed for several front-lens geometries. The adopted design incorporates one or two annular electrodes concentric with the extractor electrode. By adjusting the potentials applied to these electrodes, the electric field in front of the emitting surface can be shaped independently for the different operating modes. The same electron-optical column can therefore be switched between real-space imaging, momentum imaging, and large-area converter imaging solely by changing the lens voltages. The underlying lens design and its aberration coefficients are discussed in Ref.~\cite{tkachMultimodeObjectiveLens2025}.

\begin{figure}[ht] 
\begin{center}
\includegraphics[width=1\textwidth]{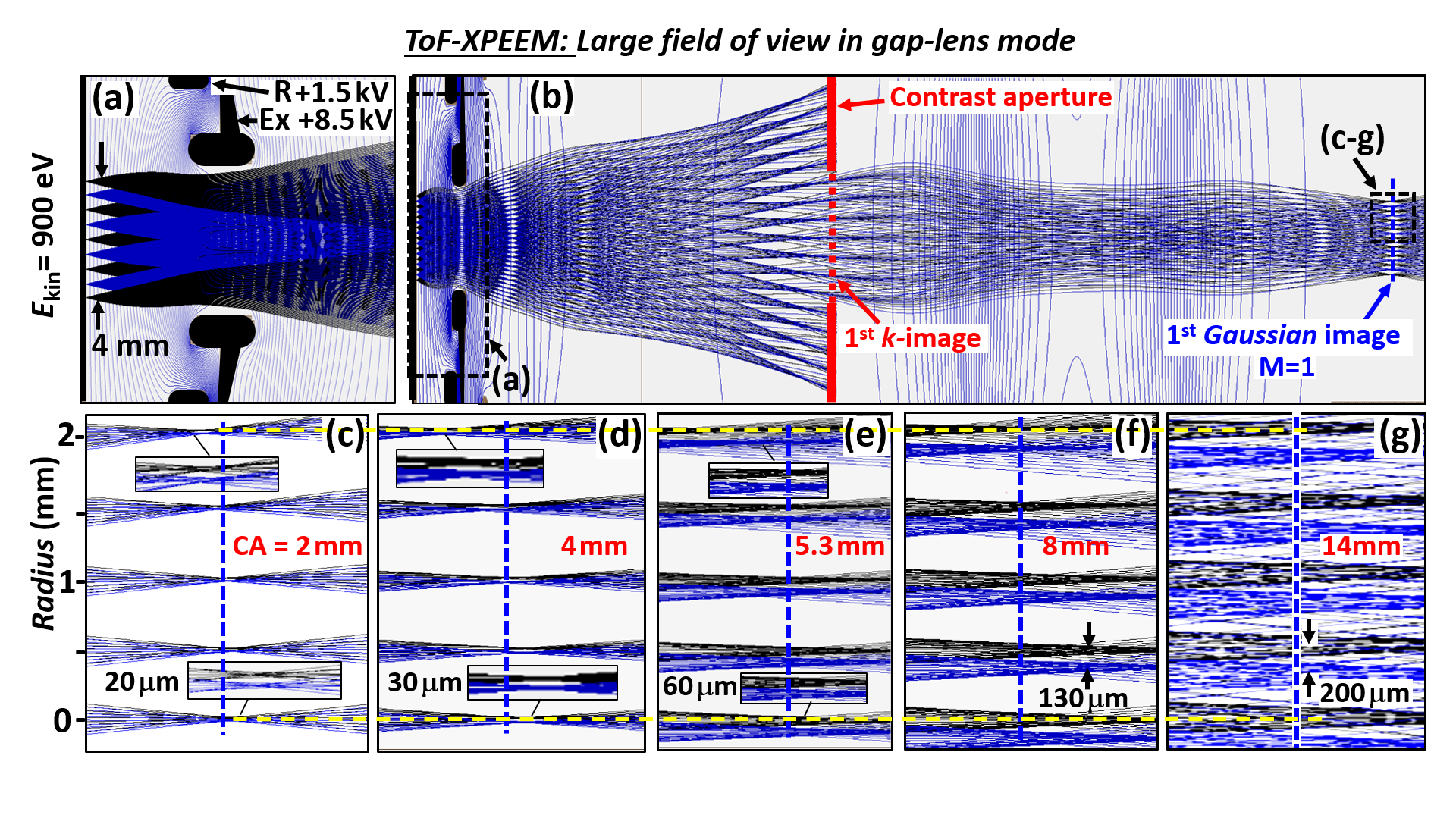} 
\end{center}
\caption{Simulations for the ToF-PAX-RIXS mode for $E_{kin} = 900$ eV and 4 mm FoV using the gap-lens mode [$U_{Ex} = 8.5$ kV, $U_R = +1.5$ kV, $F=$ 630 V/mm]. (a) Front part showing the ray bundles in the interval $\alpha_0 = -8^{\circ}$ to $8^{\circ}$ ($k$-field diameter 4.2 \AA$^{-1}$) that are contracted by the gap-lens field. (b) Rays up to the 1$^{st}$ Gaussian image at a magnification of M = 1. This low magnification of the real image requires large sizes of the contrast apertures in the reciprocal image plane. The curved stripes in (a) and (b) are moiré artifacts. (c-g) Details of the Gaussian images simulated for contrast aperture diameters between 2 and 14 mm as indicated. The panels show black and blue ray groups starting from the sample at distances $\Delta x$ between 20 $\mu$m (c) and 200 $\mu$m  (g). Reproduced from~\protect\cite{tkachMultimodeObjectiveLens2025}.}
\label{XPEEM_simulation}
\end{figure}

\subsection{RIXS configuration}
The PAX-RIXS configuration is derived from the PEEM mode lens setting, but differs in some key aspects.
Energy-filtered PEEM imaging for the X-ray region using cathode-lens instruments was first proposed by Tonner~\cite{tonnerEnergyfilteredImagingElectrostatic1990} and was subsequently demonstrated experimentally using synchrotron radiation by Bauer and coworkers~\cite{bauerSpectromicroscopyLowEnergy1997}. 
In a conventional cathode-lens geometry, the emitting
surface forms part of the electrostatic lens and is positioned only a few millimeters from the extractor electrode. The present instrument requires a substantially larger gap and, in PAX-RIXS mode, an unusually large field of view.

Figure~\ref{XPEEM_simulation} shows electron-ray-tracing calculations for the PAX-RIXS configuration at an electron kinetic energy of 900~eV and for a 4-mm-diameter converter field of view. The calculations were performed in the gap-lens mode using the voltages indicated in the figure caption. Under these conditions, the electric field generated by the annular electrode contracts the off-axis ray bundles
and directs them toward the electron-optical axis. The extractor aperture is then the principal geometrical limitation on the accessible converter area.

The gap-lens field contracts the ray bundles and pulls the outer bundles closer to the optical axis. Fig.~\ref{XPEEM_simulation}(b) shows the rays up to the 1$^{st}$ Gaussian image. The small magnification (here M=1) requires a large diameter of the 1$^{st}$ $k-$image. Therefore, this instrument requires large contrast apertures in the reciprocal image plane of up to 14 mm. The final magnification at the detector is M$_{tot}$ = 17, which allows observation of a 4.7 mm FoV on an 80 mm diameter detector. For this application, the performance criteria are a large FoV diameter (to capture a large range of momentum transfer) and the acceptance of a large solid angle, defined by the contrast aperture. To observe the $k-$dispersion in RIXS, 20-40 resolved points along the image diameter are sufficient.

The spatial resolution of the instrument in PAX-RIXS mode has been studied for various diameters of the contrast aperture (CA). Figures~\ref{XPEEM_simulation}(c-g) show the enlarged section of the Gaussian image marked by the small dashed rectangle on the right side of Fig.~\ref{XPEEM_simulation}(b). For each CA size, the spatial resolution is determined as the value of $\Delta x$, where the ray bundles are just resolved. Due to the spherical aberration, $\Delta x$ increases with increasing distance from the optical axis. This is the region from radius 0 to 2 mm of the FoV, see the scale in Fig.~\ref{XPEEM_simulation}(a, c). The patterns in Fig.~\ref{XPEEM_simulation}(c--g) were determined with contrast apertures of 2, 4, 5.3, 8, and 14 mm, respectively. The distances of the blue and black ray bundles are given in the panels.

The results for small CA diameters were used to check the resolution limit; 20 $\mu$m, as found for CA 2 mm (c), corresponds to 200 resolved points along the diameter of the FoV. For PAX-RIXS, a high count rate is mandatory, so the optimal operating point will be at large CA diameters that are just sufficient for the required lateral resolution. Figure~\ref{XPEEM_simulation}(f,g) show simulations for CA = 8 and 14 mm, yielding 130 and 200 $\mu$m, respectively, corresponding to 30 and 20 resolved points along the FoV diameter. For the 14 mm case, the diffuse background increases significantly, so this seems to be the practical limit. This value corresponds to $\alpha_0 = \pm 8^{\circ}$, which is a typical acceptance range of hemispherical analyzers. Such analyzers record the distribution along a line, whereas the ToF-PAX-RIXS approach simultaneously captures the full 2D region of interest with a radius of 2~mm.

\subsection{ARPES configuration}
The ARPES mode with a ToF momentum microscope and delay-line detector has been described in detail in Ref.~\cite{medjanikDirect3DMapping2017a} and the extension with the multimode front lens in Ref.~\cite{tkachMultimodeObjectiveLens2026}. By adjusting the potential of the ring electrode (R in Fig.~\ref{XPEEM_simulation}(a)), the electric field between the sample and the first electrode can be shaped. The extractor electrode at a high positive voltage, surrounded by the annular electrode at a lower voltage, forms a gap lens between the sample and the first electrode. This tunable lens field in the region close to the sample enables enlarged gap sizes in comparison with conventional cathode lens geometries with a homogeneous extractor field.

The field strength at the sample surface can be varied between typical values of 3 - 10 kV/mm for the extractor mode to values of 0.4 - 1.5 kV/mm in the gap-lens mode. This greatly reduces the likelihood of field emission or flashovers, as these phenomena often exhibit threshold behavior as a function of the field strength. Moreover, in the gap-lens mode, the field curvature of the $k-$image in the back focal plane is significantly reduced, and the depth of focus is increased. The ARPES configuration using the gap-lens mode has proven advantageous in previous soft X-ray photoemission experiments at beamline P04~\cite{fedchenkoObservationTimereversalSymmetry2024a, elmersChiralityKagomeMetal2025a}.

By applying a negative potential to the ring electrode R, even a retarding field can be created at the sample surface. Slow electrons are deflected back to the surface; hence, we term this the repeller mode. In this mode, the major contributor to the Coulomb forces is eliminated, and space charge-induced energy shifts and broadening are reduced. This is particularly important for experiments at X-ray FELs, where both high photoelectron energies and slow electrons released by the pump pulse are involved. A first iteration of the repeller mode (in which the extractor served as the repeller electrode) was investigated at the FEL FLASH at DESY~\cite{schonhenseSuppressionVacuumSpacecharge2021} and subsequently used in experiments~\cite{pressaccoSubpicosecondMetamagneticPhase2021, shokeenRealtimeObservationNonequilibrium2024}.

\section{Instrument performance and validation}

\subsection{Benchmarking of the PAX-RIXS mode}

To benchmark the spectral performance of the PAX-RIXS mode, measurements were performed on an infinite-layer CaCuO$_2$ (001) thin film. CaCuO$_2$ provides a suitable reference system because its Cu $L_3$-edge RIXS spectrum contains well-established low-energy magnetic excitations and higher-energy $dd$ excitations that have previously been measured with high-resolution grating spectrometers~\cite{martinelli2022}. The comparison therefore tests both the fidelity of the directly measured PAX spectra and the reliability of the reconstructed RIXS spectra.

All measurements were performed at room temperature with the incident photon energy tuned to the maximum of the Cu $L_3$ absorption resonance near 932~eV. Circularly polarized incident light was used. The incident-energy bandwidth was approximately 200~meV, and the energy resolution of the time-of-flight electron analyzer was set to approximately 150~meV. Au, Ag, and Pt converters with nominal conversion-layer thicknesses of 20~nm were measured under identical scattering and electron-optical conditions. The acquisition time was approximately 8~h for each spectrum.

\begin{figure}[ht] 
\begin{center}
\includegraphics[width=1\textwidth]{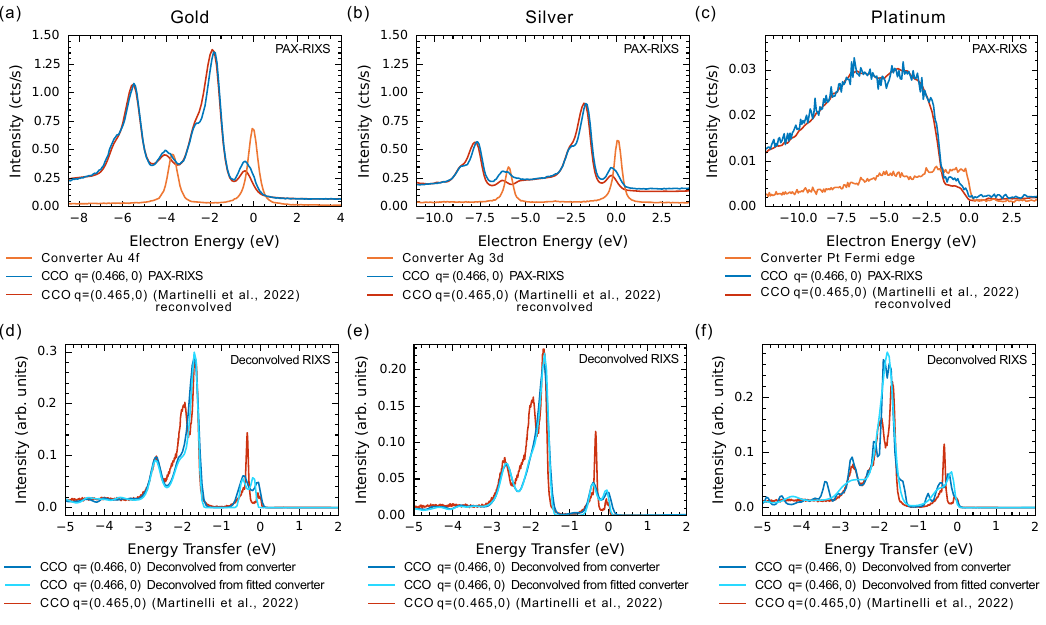} 
\end{center}
\caption{Top-row panels compare the PAX-RIXS spectra (dark blue lines) with the convoluted grating RIXS (red lines) reproduced from~\protect\cite{martinelli2022} on CCO. The orange lines represent the reference spectrum of the converter signal taken on polycrystalline graphite. Bottom-row panels show the deconvolved PAX-RIXS signal compared with grating RIXS spectra reproduced from~\protect\cite{martinelli2022}. Dark blue corresponds to the deconvolved spectra using the converter signal, while light blue corresponds to the deconvolved spectra using the fitted spectra of the converter.
From left to right, the panels correspond to Au, Ag, and Pt converters. 
The deconvolution is performed with a regularization strength of $10^{-5}$, an energy step of 50 meV, and $10^{5}$ iterations for the Au and Ag converters, and $ 5 \times 10^{5}$ iterations for the Pt converter.}
\label{Data_CCO}
\end{figure}

Figure~\ref{Data_CCO} compares the PAX-RIXS spectra obtained with the three converters and the corresponding reconstructed RIXS spectra. For this
comparison, the signal was integrated over the full illuminated region of the converter.
For the geometry shown in Fig.~\ref{RIXS_geometry}, this corresponds to a mean in-plane momentum transfer of approximately $H=0.47$~r.l.u. and to an integration interval of
$0.42 < H < 0.49$~r.l.u. and $-0.1 < L < 0.1$~r.l.u. The electron kinetic energy scale was calibrated using reference measurements on polycrystalline graphite.

The upper row of Fig.~\ref{Data_CCO} shows the directly measured
PAX-RIXS spectra together with the converter response obtained from elastically scattered photons for each of the three Au (a), Ag (b), and Pt (c) converters acquired under otherwise comparable conditions. To compare the measurements with conventional RIXS data without introducing uncertainties associated with deconvolution, the grating-RIXS spectrum reported in Ref.~\cite{martinelli2022} was convolved with the experimentally measured response of each converter. The resulting forward-calculated spectra are shown by the red curves.
This comparison provides a direct test of whether the measured PAX signal is consistent with the established RIXS spectrum and the independently measured instrumental response.

For the Au converter (Fig.~\ref{Data_CCO}(a)), the elastic-reference spectrum exhibits the Au~$4f_{7/2}$ and Au~$4f_{5/2}$ components, separated by approximately 3.6~eV, with a measured FWHM of approximately 0.45~eV. The corresponding CaCuO$_2$
PAX-RIXS signal (blue line) contains two replicas of the loss spectrum associated with the spin--orbit-split Au~$4f$ levels. The low-energy magnetic response produces a shoulder several hundred meV below each elastic feature, whereas the Cu $dd$ excitations appear at energy losses of approximately 1.5--3~eV. The forward-convolved grating-RIXS spectrum reproduces the positions and overall line shape of the measured PAX signal.
\\
For the Ag converter (Fig.~\ref{Data_CCO}(b)), the elastic-reference spectrum is dominated by the Ag~$3d_{5/2}$ and Ag~$3d_{3/2}$ core levels, separated by approximately 6~eV.
The measured linewidth of the individual components is approximately 0.36~eV.
The larger spin--orbit splitting and narrower core-level response reduce the overlap between the replicated loss spectra and provide a more favorable kernel for subsequent deconvolution. Under the present conditions, the conversion efficiency is slightly lower than that of the Au converter. The forward-convolved reference spectrum again reproduces the main features of the measured PAX-RIXS signal.
\\
The Pt converter (Fig.~\ref{Data_CCO}(c)) provides a qualitatively different response because the spectral analysis is based on the valence-band emission and Fermi edge rather than on a spin--orbit-split core level. At room temperature, the measured Fermi edge width is approximately 100~meV, smaller than both the incident photon bandwidth and the time-of-flight analyzer contribution used in the present experiment. The conversion efficiency is, however, substantially lower than for Au and Ag (approximately 40 times smaller). Consequently, the CaCuO$_2$ PAX-RIXS spectrum exhibits a considerably lower signal-to-noise ratio.
Nevertheless, an increase in intensity at energy losses of a few hundred meV and a broader response above approximately 1.5~eV are visible and are consistent with the magnetic and $dd$-excitation regions, respectively.

The lower row of Fig.~\ref{Data_CCO} (panels (d,e,f)) shows RIXS spectra reconstructed from the measured PAX signals. The deconvolution was performed using the algorithm from~\cite{higley2022}, with the experimentally measured converter response as the kernel. To evaluate the influence of statistical noise in the kernel, a second reconstruction was performed using an analytical fit to the converter response. 
In both cases, the intensities were interpolated onto an equally spaced energy grid, as required by the reconstruction algorithm~\cite{higley2022}.
Unless otherwise stated, the calculations used an energy step of 50~meV, $10^{5}$ iterations, and a regularization parameter of $10^{-5}$. These parameters were selected to obtain stable peak positions while limiting amplification of statistical noise.

For the Au converter (Fig.~\ref{Data_CCO}(d)), the reconstruction based on the directly measured kernel resolves the elastic contribution from the magnetic excitation near an energy loss of 0.4~eV and reproduces the broader $dd$-excitation manifold between approximately
1.5 and 3~eV. The reconstructed peak energies agree with the grating-RIXS data within the experimental resolution. Differences in relative intensity are expected because the present measurements used circularly polarized incident light, whereas the reference data were acquired using linear polarization. Specifically, we assume that the significantly lower intensity in the deconvolution of the second $dd$-excitation around -2~eV as compared to the grating RIXS spectrum is due to the intrinsically different response of the sample to the differently polarized excitation radiation.
Reconstruction with the fitted converter response yields a smoother spectrum but produces small shifts in some spectral features.
\\
The Ag-based reconstruction (Fig.~\ref{Data_CCO}(e)) gives the most robust result among the three converters.
The elastic and magnetic contributions are clearly separated, and their energy positions closely reproduce those of the grating-RIXS reference. The improved reconstruction is consistent with the narrower Ag~$3d$ response and the larger separation between its spin--orbit components, which reduce overlap between replicated spectral features and improve the conditioning of the inverse problem.
\\
For Pt (Fig.~\ref{Data_CCO}(f)), the reconstruction is limited primarily by the low count rate and the correspondingly poor statistical quality of both the sample and reference spectra.
The $dd$-excitation regions are qualitatively reproduced in both reconstructions using the measured and fitted kernels.  In contrast, the elastic and magnetic contributions are qualitatively recovered only with the fitted kernel, appearing as a peak with a shoulder rather than two clearly separated features. The comparison between the two reconstructions illustrates the sensitivity of the deconvolution to noise in the converter response. In particular, statistical fluctuations in the raw kernel can be amplified and generate artificial fine structure in the reconstructed spectrum. 

The effective energy resolution of the reconstructed Au- and Ag-based spectra was estimated from Gaussian fits to the reconstructed elastic feature and is approximately 220~meV under the present measurement conditions. This value is smaller than the measured width of the corresponding converter core levels because the converter response is included explicitly in the deconvolution. It should therefore be regarded as the effective resolution of the reconstructed spectrum rather than the directly measured resolution of the raw PAX signal. The uncertainty of this estimate is governed not only by the statistical error of the fit but also by the stability of the deconvolution with respect to the kernel, regularization, and iteration number.
\\

\begin{figure}[ht] 
\begin{center}
\includegraphics[width=0.95\textwidth]{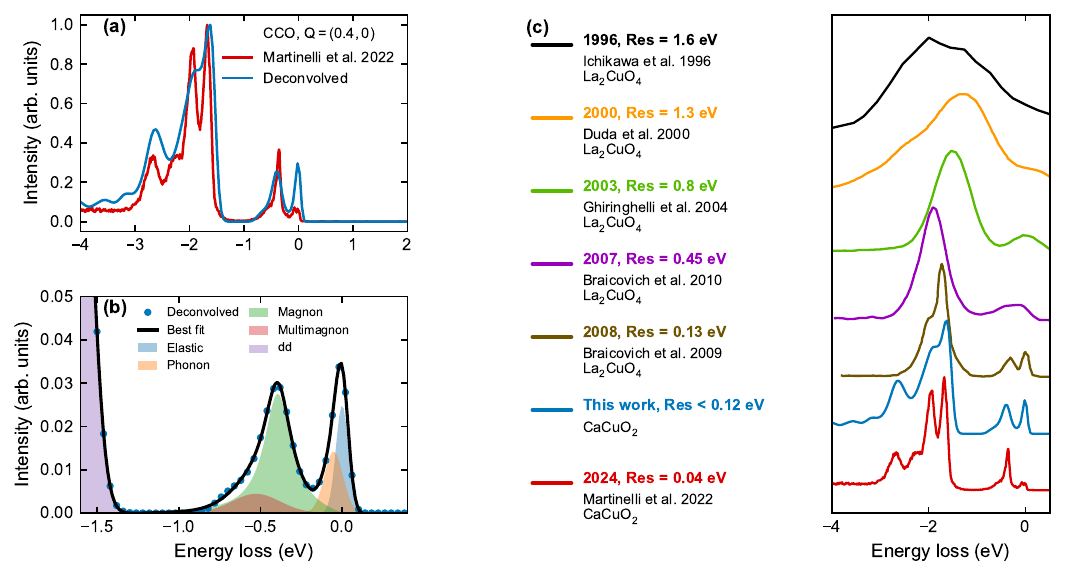} 
\end{center}
\caption{(a) Comparison of the higher-resolution deconvolved PAX-RIXS signal with the grating RIXS spectrum reproduced from~\protect\cite{martinelli2022}. The deconvolution is performed with 500 iterations, a regularization strength of $10^{-2}$, and an energy step of 40~meV. (b) Fit of the deconvolved PAX-RIXS spectrum with Gaussian functions for the elastic, phonon, multimagnon, and $dd-$excitation peaks. A damped harmonic oscillator model is used to fit the magnon peak. The elastic peak has a FWHM of 90 meV. (c) Development of the energy resolution in grating RIXS spectrometers compared to this work. Data on La$_2$CuO$_4$ were reproduced from~\protect\cite{ichikawaRXE1996, duda2000, ghiringhelli2004a, braicovich2009}, and on CaCuO$_2$ from~\protect\cite{martinelli2022}.}
\label{CCO_res}
\end{figure}

Figure~\ref{CCO_res}(a) shows a more recent measurement with improved statistics, obtained using a larger delay-line detector (with a diameter of 80~mm) with higher detection efficiency, and a silver converter. The combined instrumental and ToF energy resolution was $\simeq 180$~meV. The resulting deconvolved spectrum (in blue) reveals an asymmetric elastic line consistent with phonon excitations, as well as a magnon peak accompanied by a higher-energy shoulder attributed to multimagnon excitations.
The comparison with the measurement reported in~\cite{martinelli2022} (in red), acquired using a grating spectrometer, shows improved overall agreement. 

Figure~\ref{CCO_res}(b) shows a fit to the low-energy region of the spectrum, including contributions from the elastic line, phonon, magnon, and multimagnon excitations. When the phonon contribution is included, a Gaussian fit to the elastic component yields an estimated energy resolution of approximately 90~meV. In contrast, omitting the phonon contribution results in a resolution of approximately 120~meV, but produces a poorer overall fit. 

Figure~\ref{CCO_res}(c) places the effective resolution achieved in the present experiment in the context of the historical development of Cu $L_3$-edge grating RIXS. The present value, estimated to be better than 120~meV, remains below the resolution available at current state-of-the-art grating spectrometers, but it is sufficient to have a clear separation between the elastic and the magnetic contributions in CaCuO$_2$, while also revealing a phonon and a higher-energy multimagnon contribution. This result represents a substantial improvement over the earlier PAX-RIXS implementation reported in Ref.~\cite{Schunck2024}.
Together with the forward-convolution comparison in Fig.~\ref{Data_CCO}, this demonstrates that the present instrument reproduces the principal spectral features and energy scales measured by conventional RIXS.

\begin{figure}[ht] 
\begin{center}
\includegraphics[width=0.45\textwidth]{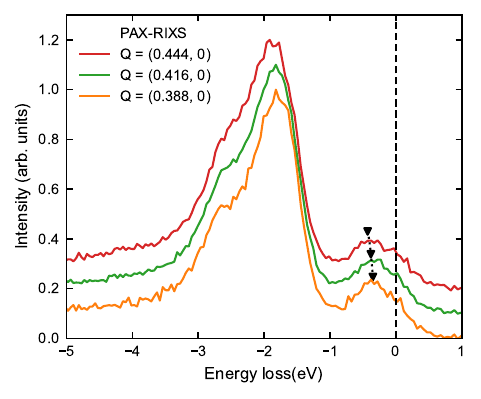} 
\end{center}
\caption{Dispersion in PAX-RIXS (non-deconvolved) spectra. The dashed line marks the elastic peak position, while the triangles and dotted lines are guides to the eye, highlighting the magnon peak's dispersive behavior.}
\label{dispersion_PAXRIXS}
\end{figure}

To study dispersion, the signal can be integrated over different detector regions that cover a finite range of transferred momentum, as discussed in section~\ref{converter_maps}. Figure~\ref{dispersion_PAXRIXS} shows an example of this approach, where the raw PAX-RIXS data are integrated over three detector regions centered around $Q=(0.388,0)$, $Q=(0.416,0)$, and $Q=(0.444,0)$~r.l.u. respectively. The elastic line is indicated by a dashed vertical line, while the magnon peak is marked by an inverted triangle.\\
With increasing momentum transfer, the magnon shifts towards higher energy loss, revealing the expected hardening of the magnon excitations. This demonstrates the ToF-PAX-RIXS instrument's ability to resolve the momentum dependence of an excitation spectrum. However, counting statistics limit the present measurements, and improved statistics are required to obtain a reliable deconvolved RIXS spectrum when integrating over smaller detector regions.

Further improvements require increased count rates and a sharper converter response.
The Pt converter can be cooled independently of the sample to approximately 30~K, reducing the thermal contribution to the Fermi edge width to the order of 10~meV.
This value represents the converter-related thermal broadening only; the total achievable RIXS resolution will additionally depend on the incident photon bandwidth, the time-of-flight analyzer resolution, counting statistics, calibration stability, and the robustness of the deconvolution. Owing to the low Pt conversion efficiency, substantially longer acquisition times or increased detection efficiency will be required to exploit this sharper response.

\subsection{Combined ARPES and PAX-RIXS measurements}
The combined ARPES and PAX-RIXS operation of the instrument was demonstrated using a cleaved Ca-doped Y$_{0.89}$Ca$_{0.11}$Ba$_2$Cu$_3$O$_{6.45}$ single crystal, hereafter denoted Ca-YBCO.
The same mounted sample was measured sequentially in momentum-microscopy ARPES and PAX-RIXS modes without transfer to a separate endstation. The sample was cleaved \textit{in situ} with the crystallographic $c$ axis normal to the exposed surface,
and all measurements were performed at 30~K. The energy resolution in ARPES mode was set to 36~meV.

\begin{figure}[!ht]
\begin{center}
\includegraphics[width=1\textwidth]{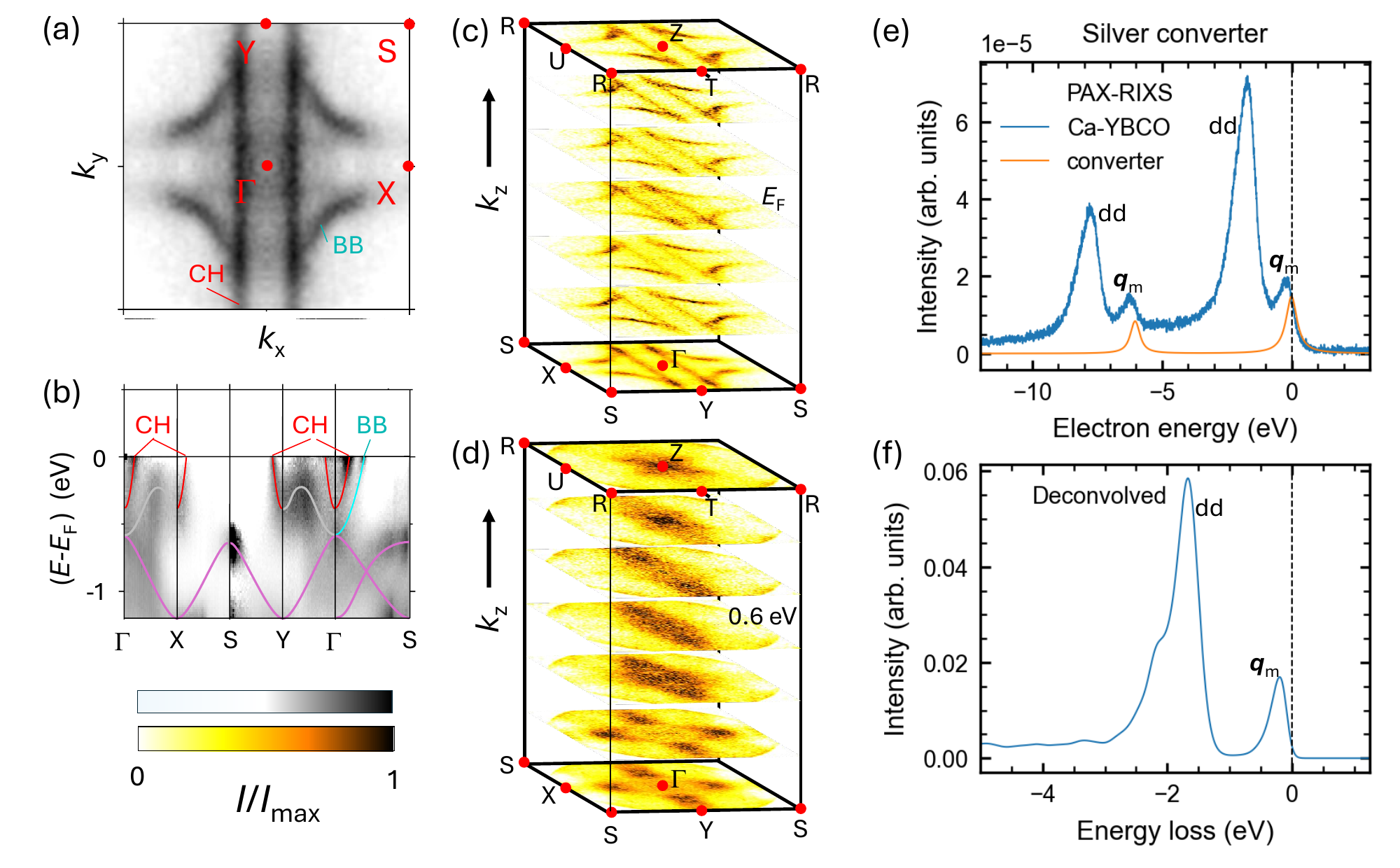} 
\end{center}
\caption{
Combination of ToF-ARPES and RIXS on a cleaved Ca-YBCO surface. 
(a) Photoemission intensity distribution $I(E_F,k_x,k_y)$ of the Fermi surface section in the $\Gamma$-X-Y plane.
The Fermi surface section shows the CuO chains (CH) and the bonding bands of the CuO$_2$ planes (BB).
(b) Sections of the photoemission intensity, $I(E_B,k_x,k_y)$,  along the high-symmetry directions reveal the band dispersion in the valence band range.
(c,d)  Constant energy sections $I(E_B,k_x,k_y,k_z)$ at the indicated binding energies. The photon energy in the range from $h\nu=275-305$~eV determines the perpendicular momentum $k_z$ using the free-electron final-state model. 
(e) PAX-RIXS spectra of X-rays scattered from a carbon tape (orange) and from Ca-YBCO (blue) in the back-scattering geometry (transferred momentum H$=0.34$~r.l.u.). The energy scale has been shifted to set the Ag $3d$ core-level peak for elastically scattered photons at zero.
(f) The deconvolved PAX-RIXS signal shows magnon-induced energy losses at low energy, as well as energy losses caused by $dd$ excitations within the energy range of $\Delta E_{\rm Loss}=-1$ to -3~eV. The magnon peaks for momenta around H$=0.34$~r.l.u. show an energy loss $E(q_m)\approx 200-300$ meV.   
}
\label{Data_YBCO}
\end{figure}

In momentum-microscopy mode, the instrument records the photoemission intensity $I(E_{\mathrm{B}},k_x,k_y)$ over the full two-dimensional momentum field of view.
The perpendicular momentum component was varied by changing the photon energy. Within a free-electron final-state approximation~\cite{medjanikDirect3DMapping2017a}, photon energies of 270 and 305~eV correspond to approximately $k_z=8.0G_{001}$ and $8.5G_{001}$, respectively, where $G_{001}$ is the reciprocal-lattice vector along the $\Gamma$--$Z$ direction.

Figure~\ref{Data_YBCO}(a) shows the Fermi-surface section in the
$\Gamma$--$X$--$Y$ plane. The observed twofold symmetry from the orthorhombic crystal structure indicates that the measurement predominantly probes a single structural domain. The quasi-one-dimensional Cu--O chain states and the bonding bands derived from the CuO$_2$ planes are visible in the momentum map and in the energy--momentum sections shown in Fig.~\ref{Data_YBCO}(b).

The photon-energy scan shown in Figs.~\ref{Data_YBCO}(c) and (d) demonstrates the capability to acquire four-dimensional photoemission data, $I(E_{\mathrm{B}},k_x,k_y,k_z)$. The Fermi-level intensity exhibits only weak $k_z$ dependence, as expected for the quasi-two-dimensional CuO$_2$-derived states, whereas deeper valence-band features show a more pronounced variation with perpendicular momentum.

Figure~\ref{Data_YBCO}(a) shows a cross-section of the three-dimensional Fermi surface in the $\Gamma$-X-Y plane. The twofold symmetry of the pattern reflects the twofold orthorhombic symmetry perpendicular to the $c$-axis. 
The deviation from fourfold symmetry indicates that a single-domain area was probed in this experiment.
The pair of straight lines (CH) in the direction of the $\Gamma$-Y axis indicates the Fermi wave vectors of the bands induced by the Cu-O chains oriented along the horizontal axis in direct space. 
These bands exhibit shallow dispersion perpendicular to the chain direction but pronounced electron-like dispersion along the chain ([)see also the band dispersion in the cross sections shown in Fig.~\ref{Data_YBCO}(b)).

After completion of the ARPES measurement, the instrument was switched to PAX-RIXS mode by inserting the Ag converter and changing the electron-optical lens voltages to real-space imaging mode. Figure~\ref{Data_YBCO}(e) compares the Ca-YBCO PAX-RIXS spectrum with an elastic-reference spectrum obtained from carbon tape under the same instrumental conditions. The time-of-flight analyzer contribution to the energy resolution was set to approximately 125~meV.

Relative to the elastic-reference spectrum, the Ca-YBCO signal shows additional spectral weight at energy losses of a few hundred meV and between approximately 1 and 3~eV. Following deconvolution, Fig.~\ref{Data_YBCO}(f) reveals a low-energy magnon feature at approximately 200--300~meV and the broader Cu $dd$-excitation manifold. For a mean in-plane momentum transfer of $H\approx0.34$~r.l.u., the magnon energy scale agrees with previous Cu $L_3$-edge RIXS measurements on YBCO-related compounds~\cite{cava1987, magnuson2014, minola2015}.

Because different positions on the converter correspond to different scattering angles, spatially resolved analysis of the converter image provides access to the momentum dependence of the loss spectrum. In the present data set, the available statistics permit extraction of a spectrum centered around $H\approx0.34$~r.l.u.

\section{Conclusion}
We have presented ToF-PAX-RIXS, a compact instrument that combines time-of-flight momentum-microscopy ARPES and PAX-based RIXS within a single electron-optical platform. Switching between the two operating modes requires only inserting or retracting a converter and adjusting the electron-optical lens voltages, allowing both measurements to be performed on the same mounted sample without transfer to a separate spectrometer.

The PAX-RIXS performance was benchmarked at the Cu $L_3$ edge using CaCuO$_2$. Measurements with Au, Ag, and Pt converters reproduce the principal spectral features observed with a conventional grating RIXS spectrometer. Forward convolution of established grating-RIXS spectra with the experimentally measured converter response shows good agreement with the directly measured PAX signals. Deconvolution further recovers clear elastic, magnetic, and $dd$-excitation signals, as well as phonon and multimagnon contributions, with an effective reconstructed energy resolution better than 120~meV under the present experimental conditions. Among the converter materials investigated, Ag provides the most favorable compromise between conversion efficiency, spectral width, and reconstruction stability.

The combined operation of the instrument was demonstrated on a cleaved Ca-doped YBa$_2$Cu$_3$O$_{6.45}$ crystal. Momentum-resolved ARPES and PAX-RIXS spectra were acquired sequentially on the same mounted sample, demonstrating the capability to correlate the electronic structure with collective excitations using a common experimental platform. In addition, real-space imaging of the converter provides simultaneous access to a finite range of scattering angles and momentum transfers.

The present implementation is primarily limited by photon-to-electron conversion efficiency, counting statistics, and the stability of the numerical reconstruction. Improved signal-to-noise ratio should enable more reliable analysis of spatially resolved converter regions and, therefore, more detailed measurements of excitation dispersions.
X-PEEM imaging of the sample gives spectroscopic information on the distribution of atomic species on the surface. This includes their chemical state, exploiting core-level shifts.

Further improvements may be achieved using a cryogenically cooled Pt converter, whose Fermi edge provides a substantially sharper spectral response than the Au and Ag core levels. At a converter temperature of approximately 30~K, the thermal contribution to the Fermi edge width is expected to approach 10~meV. The total achievable RIXS resolution will, however, also depend on the incident photon bandwidth, the time-of-flight analyzer resolution, counting statistics, calibration stability, and the robustness of the deconvolution procedure. Ongoing developments therefore focus on increasing detection efficiency and the transmission of the microscope optics, and improving the reliability of spectral reconstruction. These results establish ToF-PAX-RIXS as a new approach to compact, multimodal soft-X-ray spectroscopy, providing complementary access to single-particle electronic structure and collective excitations within a common instrument.

\begin{acknowledgements}
We thank Giacomo Ghiringhelli for providing us with the CaCuO$_2$ RIXS data. We thank Dirk Zimmer, Samuel Culeca, Ivan Braining, and Jan Torben R\"oh for technical assistance during instrument installations and the DESY mechanical engineering team for the design and construction of a positioning stage. 
We thank Christian Budzylek from Cryoandmore for his efforts in commissioning the closed-cycle cooling system. 
We thank Harshit Agarwal and Luis Filsinger for help and discussions during the beamtime.
We also thank DESY (Hamburg, Germany) for providing the experimental facilities. This research was carried out at PETRA III at beamline P04. 
\end{acknowledgements}

\begin{funding}
This work was funded by the Deutsche Forschungsgemeinschaft (DFG, German Research Foundation), grant no. TRR288–422213477 (project B03 and B04).
We gratefully acknowledge funding for the instrument from the Federal Ministry of Education and Research (BMFTR) under the framework program ErUM-Pro (projects 05K25CK1, 05K25UM1, 05K22CK1, and 05K22UM4). T.L. acknowledges funding from the European Research Council (ERC) under the European Union’s Horizon 2020 and Horizon Europe research and innovation program through the ERC Synergy Grant HERO (Grant Agreement No. 810451) and under the Marie Skłodowska-Curie grant agreement No. 101274202. J.O.S. acknowledges support from the Wallenberg Initiative Materials Science for Sustainability (WISE), funded by the Knut and Alice Wallenberg Foundation.
\end{funding}

\ConflictsOfInterest{The authors declare no conflicts of interest.}

\DataAvailability{Data are available upon request.}

\bibliography{references,addreferences}

\end{document}